\documentclass{llncs}
\usepackage{amssymb,amsmath}
\usepackage{gastex}
\usepackage{epsfig}
\usepackage{times}

\newcounter{compressEnum}

{}

\newtheorem{mydefi}{Definition}

\newenvironment{myProof}[1][\unskip]{\smallskip\par\noindent{\bfseries Proof
    #1.}\ \ 
        \global\def\qed{\origQED\global\def\qed{}}\penalty10000}%
        {\qed\par\smallskip\global\def\qed{\origQED\global\def\qed{}}}

\def\endof{%
  \leavevmode
  \parfillskip=0pt%
  \widowpenalty=10000%
  \displaywidowpenalty=10000%
  \finalhyphendemerits=0%
  \unskip\nobreak\null\hfil\penalty50\hskip2em\null\hfill%
}
\def\eodsymbol{\ensuremath\square}
\def\eopsymbol{\ensuremath\blacksquare}

\def\origEOD{\nobreak\leavevmode\endof\eodsymbol\par}
\def\EOD{\origEOD\global\def\EOD{}}
\def\origQED{{\nobreak\leavevmode\endof\eopsymbol\par\smallskip}}
\def\qed{\origQED\global\def\qed{}}

\newcommand{\Pref}{{\sf Prefs}}
\newcommand{\Play}{{\sf Plays}}
\newcommand{\K}{{\sf K}}
\newcommand{\Last}{{\sf Last}}

\newcommand{\POG}{{\sf POG}}
\newcommand{\POMDP}{{\sf POMDP}}

\def\abs#1{\ensuremath{\lvert #1\rvert}}

 \renewcommand{\L}{{\mathcal{L}}}

\newcommand{\reach}{\mathsf{Reach}}
\newcommand{\Bad}{\mathsf{Bad}}

\newcommand{\outcome}{\mathsf{outcome}}
\newcommand{\safe}{\mathsf{Safe}}

\newcommand{\Outcome}{{\mathsf{Outcome}}}
\newcommand{\trans}{{\delta}}

\newcommand{\real}{{\mathbb R}}

\renewcommand{\l}{{\ell}}

\newcommand{\nat}{\mathbb N} 
\newcommand{\tuple}[1]{\langle #1 \rangle}

\newcommand{\Post}{\mathsf{Post}}

\newcommand{\Obs}{{\cal{O}}}

\newcommand{\obs}{\mathsf{obs}}

\newcommand{\target}{{\cal T}}

\newcommand{\straa}{\alpha}
\newcommand{\strab}{\beta}
\newcommand{\dist}{{\cal D}}
\newcommand{\Supp}{{\sf Supp}}

\newcommand{\Straa}{{\cal A}}
\newcommand{\Strab}{{\cal B}}

\newcommand{\Reach}{\mathsf{Reach}}
\newcommand{\Until}{\mathsf{Until}}
\newcommand{\Buchi}{{\sf B\ddot{u}chi}}
\newcommand{\coBuchi}{\mathsf{coB\ddot{u}chi}}
\newcommand{\Parity}{\mathsf{Parity}}

\newcommand{\supp}{\Supp}

\newcommand{\Safe}{\mathsf{Safe}}

\newcommand{\Prb}{\mathrm{Pr}}
\newcommand{\set}[1]{\{\: #1 \:\}}

\newcommand{\Inf}{\mathsf{Inf}}
\newcommand{\Nats}{\mathbb{N}}

\newcommand{\ov}{\overline}

\newcommand{\tick}{{\sf tick}}
\newcommand{\Goal}{\mathsf{Goal}}

\makeatletter

\begingroup \catcode `|=0 \catcode `[= 1
\catcode`]=2 \catcode `\{=12 \catcode `\}=12
\catcode`\\=12 |gdef|@xcomment#1\end{comment}[|end[comment]]
|endgroup

\def\@comment{\let\do\@makeother \dospecials\catcode`\^^M=10\def\par{}}

\def\begincomment{\@comment\@xcomment}

\makeatother

\newenvironment{comment}{\begincomment}{}

\begin{document}

\title{{\bf Qualitative Analysis of Partially-observable Markov 
Decision Processes}}

\author{Krishnendu Chatterjee\inst{1} \and Laurent Doyen\inst{2} \and  Thomas A. Henzinger\inst{1} }

\institute{
IST Austria (Institute of Science and Technology Austria) \\
\and LSV, ENS Cachan \& CNRS, France 
}

\maketitle
\thispagestyle{empty}

\begin{abstract}
We study observation-based strategies for \emph{partially-observable Markov 
decision processes} (\POMDP s) with parity objectives. 
An observation-based strategy relies on partial information about the history 
of a play, namely, on the past sequence of observations.
We consider qualitative analysis problems: given a \POMDP\/ with a 
parity objective, decide whether there exists an observation-based strategy to
achieve the objective with probability~1 (almost-sure winning), 
or with positive probability (positive winning).
Our main results are twofold. 
First, we present a complete picture of the computational complexity of
the qualitative analysis problem for \POMDP s with parity objectives and 
its subclasses: safety, reachability, B\"uchi, and coB\"uchi objectives.
We establish several upper and lower bounds that were not known in the literature, and
present efficient and symbolic algorithms for the decidable subclasses.
Second, we give, for the first time, optimal bounds (matching upper and lower bounds) for the 
memory required by pure and randomized observation-based strategies for all 
classes of objectives. 
\end{abstract}

\section{Introduction}

\noindent{\bf Markov decision processes.} A \emph{Markov decision process
(MDP)} is a model for systems that exhibit both probabilistic and 
nondeterministic behavior.
MDPs have been used to model and solve control problems 
for stochastic systems: there, nondeterminism represents the freedom of the controller 
to choose a control action, while the probabilistic component of the behavior 
describes the system response to control actions. 
MDPs have also been adopted as models for concurrent probabilistic systems, 
probabilistic systems operating in open environments~\cite{SegalaT}, and 
under-specified probabilistic systems \cite{BdA95}.

\smallskip\noindent{\bf System specifications.} 
The \emph{specification} describes the set of desired behaviors of
the system, 
and is typically an $\omega$-regular set of paths. 
Parity objectives are a canonical way to define such specifications in MDPs.   
They include reachability, safety, B\"uchi 
and coB\"uchi 
objectives as special cases.
Thus MDPs with parity objectives provide the theoretical framework to 
study problems such as the verification 
and the control of stochastic systems.

\smallskip\noindent{\bf Perfect vs. partial observations.} 
Most results about MDPs make the hypothesis of \emph{perfect observation}. 
In this setting, the controller always knows, while interacting with the system (or MDP), 
the exact state of the MDP. 
In practice, this hypothesis is often unrealistic. 
For example, in the control of multiple processes, each process has only access 
to the public variables of the other processes, but not to their private variables.
In the control of hybrid systems~\cite{BouyerDMP03,DDR06}, or in automated planning~\cite{MadaniHC03}, 
the controller usually has noisy information about the state of the systems due to finite-precision sensors.
In such applications, MDPs with \emph{partial observation} (POMDPs) provide a more appropriate model.

\smallskip\noindent{\bf Qualitative and quantitative analysis.} 
Given an MDP with parity objective, the \emph{qualitative analysis} 
asks for the computation of the set of \emph{almost-sure winning} states
(resp., \emph{positive winning} states) in which the controller can 
achieve the parity objective with probability~1 (resp., positive probability);
the more general \emph{quantitative analysis} asks for the computation 
at each state of the maximal probability with which the controller can satisfy the 
parity objective. 
The analysis of POMDPs is considerably more complicated than the analysis of MDPs. 
First, the decision problems for POMDPs usually lie in higher
complexity classes than their perfect-observation counterparts: for example,
the quantitative analysis of POMDPs with reachability and safety objectives is 
undecidable~\cite{Paz-Book}, whereas for MDPs with perfect observation, this question can be 
solved in polynomial time~\cite{luca-thesis,CJH04}.
Second, in the context of POMDPs, witness winning strategies for the controller need memory 
even for the simple objectives of safety and reachability. 
This is again in contrast to the perfect-observation case, where memoryless 
strategies suffice for all parity objectives.
Since the quantitative analysis of POMDPs is undecidable (even for computing 
approximations of the maximal probabilities~\cite{MadaniHC03}), 
we study the qualitative analysis of POMDPs with parity objective and its subclasses. 

\smallskip\noindent{\bf Contribution.}
For the qualitative analysis of POMDPs, the following results are known:
(a)~the problems of deciding if a state is almost-sure winning for 
reachability and B\"uchi objectives can be solved in EXPTIME~\cite{BBG08};
(b)~the problems for almost-sure winning for coB\"uchi objectives and 
positive winning for B\"uchi objectives are undecidable~\cite{BBG08,GIMBERT:2009:HAL-00403463:3}; and
(c)~the EXPTIME-completeness of almost-sure winning for 
safety objectives follows from the results on games with partial observation~\cite{CDHR07,BD08}.
Our new contributions are as follows:
\begin{enumerate}
\item  First, we show that
(a)~positive winning for reachability objectives is NLOGSPACE-complete;
and (b)~almost-sure winning for reachability and B\"uchi objectives, and positive winning for safety and 
coB\"uchi objectives are EXPTIME-hard\footnote{A very brief (two line) 
proof of EXPTIME-hardness is sketched in~\cite{dA99} (see the discussion before Theorem~\ref{thrm_complexity} 
for more details).\label{pos-safety-hard}}.
We also present a new proof that positive winning for safety and coB\"uchi objectives can be 
solved in EXPTIME\footnote{A different proof  that positive safety can be solved in EXPTIME is given in~\cite{GriponS09} (see the discussion after Theorem~\ref{thrm_pos_safe} for a comparison).}.  
It follows that almost-sure  winning for reachability and B\"uchi, 
and positive winning for safety and coB\"uchi, are EXPTIME-complete.
This completes the picture for the complexity of the qualitative analysis for POMDPs 
with parity objectives.
Moreover our new proofs of EXPTIME upper-bound proofs yield efficient and symbolic algorithms 
to solve positive winning for safety and coB\"uchi objectives in POMDPs.

\item Second, we present a complete characterization of the amount of memory required by 
pure (deterministic) and randomized strategies for the qualitative 
analysis of POMDPs.    
For the first time, we present optimal memory bounds (matching upper and lower 
bounds) for pure and randomized strategies: we show that 
(a)~for positive winning of reachability objectives, randomized memoryless 
strategies suffice, while for pure strategies linear memory is necessary and 
sufficient; (b)~for almost-sure winning of safety, reachability, and 
B\"uchi objectives, and for positive winning of safety and coB\"uchi 
objectives, 
exponential memory is necessary and sufficient for both pure and 
randomized strategies.

\end{enumerate}

\smallskip\noindent{\bf Related work.} Though MDPs have been widely 
studied under the hypothesis of perfect observations, there are a few works 
that consider POMDPs, e.g.,~\cite{PapaTsi,Littman-thesis} for several 
finite-horizon quantitative objectives.
The results of~\cite{BBG08} shows the upper bounds for almost-sure winning for 
reachability and B\"uchi objectives,
and the work of~\cite{CSV09} considers a subclass of \POMDP s with B\"uchi objectives 
and presents a PSPACE upper bound for the subclass.
The undecidability of almost-sure winning for coB\"uchi and positive winning for 
B\"uchi objectives is established by~\cite{BBG08,GIMBERT:2009:HAL-00403463:3}.
We present a solution to the remaining problems related to the qualitative 
analysis of POMDPs with parity objectives, and complete the picture.
Partial information has been studied in the context of two-player games~\cite{Reif84,CDHR07}, 
a model that is incomparable to MDPs, though some techniques (like the subset construction) can
be adapted to the context of POMDPs. More general models of stochastic games with partial information 
have been studied in~\cite{BGG09,GriponS09}, and lie in higher complexity classes. 
For example, a result of~\cite{BGG09} shows
that the decision problem for positive winning of safety objectives is 2EXPTIME-complete in the general model,
while for POMDPs, we show that the same problem is EXPTIME-complete.

\section{Definitions}\label{sec:definitions}

A \emph{probability distribution} on a finite set $A$ is a function
$\kappa: A \to [0,1]$ such that $\sum_{a \in A} \kappa(a) = 1$. 
The \emph{support} of $\kappa$ is the set $\Supp(\kappa) = \{a \in A \mid \kappa(a) > 0\}$.
We denote by $\dist(A)$ the set of probability distributions on $A$.

\smallskip\noindent{\em Games and MDPs.}
A \emph{two-player game structure} or a \emph{Markov decision process (MDP)} 
(\emph{of partial observation}) is a tuple $G=\tuple{L,\Sigma,\trans,\Obs}$, where 
$L$ is a finite set of states, 
$\Sigma$ is a finite set of actions, 
$\Obs \subseteq 2^L$ is a set of observations that partition\footnote{A slightly more general model with 
overlapping observations can be reduced in polynomial time to partitioning observations~\cite{CDHR07}.} the state space $L$. 
We denote by $\obs(\l)$ the unique observation $o \in \Obs$ such that $\l \in o$.
In the case of games, $\trans \subseteq L \times \Sigma \times L$ is a set of labeled transitions; 
in the case of MDPs, $\trans: L \times \Sigma \to \dist(L)$ is a probabilistic transition function.
For games, we require that for all $\l \in L$ and all $\sigma \in \Sigma$, there exists 
$\l' \in L$ such that $(\l, \sigma, \l') \in \trans$.  
We refer to a game of partial observation as a \POG\/ and 
to an MDP of partial observation as a \POMDP.
We say that $G$  is a game or MDP of \emph{perfect observation}
if $\Obs = \{ \{\l\} \mid \l \in L \}$.
For $\sigma \in \Sigma$ and $s \subseteq L$,
define $\Post^G_\sigma(s) = \{\l' \in L \mid \exists \l \in s: (\l,\sigma,\l') \in \trans \}$
when $G$ is a game, and $\Post^G_\sigma(s) = \{\l' \in L \mid \exists \l \in s: \trans(\l,\sigma)(\l') >0 \}$
when $G$ is an MDP.

\medskip\noindent{\em Plays.}
Games are played in rounds in which Player~$1$ chooses
an action in $\Sigma$, and Player~$2$ resolves
nondeterminism by choosing the successor state;
in MDPs the successor state is chosen according to the probabilistic 
transition function.
A \emph{play} in $G$ is an infinite sequence 
$\pi=\l_0 \sigma_0 \l_1 \ldots \sigma_{n-1} \l_n \sigma_n \ldots$ such that 
$\l_{i+1} \in \Post^G_{\sigma_i}(\{\l_i\})$ for all $i \geq 0$.
%
%
%
%
The infinite sequence 
$\obs(\pi)=\obs(\l_0) \sigma_0 \obs(\l_1) \ldots \sigma_{n-1} \obs(\l_n) \sigma_{n} \ldots$
is the \emph{observation} of $\pi$.

%
The set of infinite plays in $G$ is denoted $\Play(G)$,
and the set of finite prefixes $\l_0 \sigma_0 \ldots \sigma_{n-1} \l_n$ of plays is denoted $\Pref(G)$.
A state $\l \in L$ is \emph{reachable} in $G$ if there exists a prefix 
$\rho \in \Pref(G)$ such that $\Last(\rho) = \l$ where $\Last(\rho)$
is the last state of $\rho$.
%
%

\begin{comment}
\begin{lemma}\label{lem:knowledge}
Let $G=\tuple{L,\Sigma,\trans,\Obs}$
be a \POG\/ or a \POMDP. 
For $\sigma \in \Sigma$,  $\l \in L$, and $\rho,\rho' \in \Pref(G)$ with 
$\rho' = \rho \cdot \sigma \cdot \l$, 
Then $\K(\obs(\rho')) = \Post^G_\sigma(\K(\obs(\rho))) \cap \obs(\l)$.
\end{lemma}
\end{comment}

\medskip\noindent{\em Strategies.}
A \emph{pure strategy} in $G$ for Player~$1$ is a function 
$\straa:\Pref(G) \to \Sigma$. 
A \emph{randomized strategy} in $G$ for Player~$1$ is a function 
$\straa:\Pref(G) \to \dist(\Sigma)$. 
A (pure or randomized) strategy $\straa$ for Player~$1$ is 
\emph{observation-based} if for all prefixes $\rho,\rho' \in \Pref(G)$, 
if $\obs(\rho) = \obs(\rho')$, then $\straa(\rho)=\straa(\rho')$. 
In the sequel, we are interested in the existence of observation-based
strategies for Player~$1$.
A \emph{pure strategy} in $G$ for Player~$2$ is a function 
$\strab:\Pref(G) \times \Sigma \to L$
such that for all $\rho \in \Pref(G)$ and all $\sigma \in \Sigma$, we have
$(\Last(\rho), \sigma, \strab(\rho,\sigma)) \in \trans$.
A \emph{randomized strategy} in $G$ for Player~$2$ is a function 
$\strab:\Pref(G) \times \Sigma \to \dist(L)$
such that for all $\rho \in \Pref(G)$, all $\sigma \in \Sigma$, and all 
$\l \in \Supp(\strab(\rho,\sigma))$, we have 
$(\Last(\rho), \sigma, \l) \in \trans$.
We denote by $\Straa_G$, $\Straa_G^O$, and $\Strab_G$ the set of all 
Player-$1$ strategies, the set of all observation-based Player-$1$ 
strategies, and the set of all Player-$2$ strategies in $G$, respectively.

\newcommand{\mem}{{\sf Mem}}

\smallskip\noindent{\em Memory requirement of strategies.} 
An equivalent definition of strategies is as follows.
Let $\mem$ be a set called \emph{memory}.
An observation-based strategy with memory can be described by two functions,
a \emph{memory-update} function $\straa_{u}$: $\mem \times \Obs \times \Sigma \to \mem$ 
that given the current memory, observation and the action updates the memory, 
and a \emph{next-action} function $\straa_{n}$: $\mem \times \Obs \to \dist(\Sigma)$
that given the current memory and current observation specifies the probability 
distribution\footnote{For a pure strategy, the next-action
function specifies a single action rather than a probability distribution.} 
of the next action, respectively.
A strategy is \emph{finite-memory} if the memory $\mem$ is finite
and the size of a finite-memory strategy $\straa$ is the size $\abs{\mem}$ of its memory.
A strategy is \emph{memoryless} if $\abs{\mem} = 1$.
The memoryless strategies do not depend on the history of a play, 
but only on the current state. 
Memoryless strategies for player~1 can be viewed as functions
$\straa$: $\Obs \to \dist(\Sigma)$.

\medskip\noindent{\em Objectives.}
An \emph{objective} for $G$ is a set $\phi$ of infinite sequences of states and actions, 
that is, $\phi \subseteq (L \times \Sigma)^\omega$. 
We consider objectives that are Borel measurable, i.e., sets in the Cantor topology 
on $(L \times \Sigma)^\omega$~\cite{Kechris}.
We specifically consider reachability, safety,
B\"uchi, coB\"uchi, and parity objectives, all of them being Borel measurable.
The parity objectives are a canonical form to express all $\omega$-regular 
objectives~\cite{Thomas97}.
For a play $\pi=\l_0\sigma_0\l_1\dots$, we denote 
by $\Inf(\pi) = \{ \l \in L \mid \l = \l_i \mbox{ for infinitely many } i\mbox{'s} \}$ 
the set of states that appear infinitely often in $\pi$.

\begin{itemize}
 	\item \emph{Reachability and safety objectives.}
	Given a set $\target \subseteq L$ of target states, the \emph{reachability} objective 
	$\Reach(\target) = \set{ \l_0 \sigma_0 \l_1 \sigma_1 \ldots \in \Play(G) \mid \exists k \geq 0:  \l_k \in \target}$
	requires that a target state in $\target$ be visited at least once.
	Dually, the \emph{safety} objective $\Safe(\target) = \set{ \l_0 \sigma_0 \l_1 \sigma_1 \ldots \in \Play(G) \mid 
	\forall k \geq 0:  \l_k \in \target}$ 
	requires that only states in $\target$ be visited; the objective $\Until(\target_1,\target_2) = 
	\{ \l_0 \sigma_0 \l_1 \sigma_1 \ldots \in \Play(G) \mid \exists k \geq 0:  \l_k \in \target_2 \land \forall j \leq k: \l_j \in \target_1\}$
	requires that only states in $\target_1$ be visited before a state in $\target_2$ is visited;

	\item \emph{B\"uchi and coB\"uchi objectives.}
	The \emph{B\"uchi} objective $\Buchi(\target) = \set{ \pi \mid \Inf(\pi) \cap \target \neq \emptyset}$ 
	requires that a state 
	in $\target$ be visited infinitely often. 
	Dually, the \emph{coB\"uchi} objective $\coBuchi(\target) =\set{\pi \mid \Inf(\pi) \subseteq \target}$ 
	requires that only states 
	in $\target$ be visited infinitely often; and
	
	\item\emph{Parity objectives.} 
	For $d \in \Nats$, let $p:L \to \set{0,1,\ldots,d}$ be a 
	\emph{priority function} that maps each state 
	to a nonnegative integer priority.
	The \emph{parity} objective $\Parity(p) = \set{\pi \mid \min\set{ p(\l) \mid \l \in \Inf(\pi)} 
	\text{ is even} }$ requires that the smallest priority that appears infinitely often be even.
\end{itemize}

Note that the objectives $\Buchi(\target)$ and $\coBuchi(\target)$ 
are special cases of parity objectives defined by respective priority functions
$p_1,p_2$ such that $p_1(\l) = 0$ and $p_2(\l) = 2$ if $\l \in \target$,
and $p_1(\l) = p_2(\l) = 1$ otherwise. 
An objective $\phi$ is \emph{visible} if it depends only on the observations;
formally, $\phi$ is visible if, whenever $\pi \in \phi$ 
and $\obs(\pi)=\obs(\pi')$, then $\pi'\in \phi$.
In this work, all our upper bound results are for the general parity objectives 
(not necessarily visible), and all the lower bound results for \POMDP s are for the 
special case of visible objectives (and hence the lower bounds also hold for general
objectives).

\medskip\noindent{\em Almost-sure and positive winning.}
An \emph{event} is a measurable set of plays, and 
given strategies $\straa$ and $\strab$ for the two players (resp., a strategy
$\straa$ for Player~1 in MDPs), the probabilities of events are uniquely defined~\cite{Var85}. 
For a Borel objective~$\phi$, we denote by $\Prb_{\l}^{\straa,\strab}(\phi)$ 
(resp., $\Prb_{\l}^{\straa}(\phi)$ for MDPs) the probability that $\phi$ is 
satisfied from the starting state~$\l$ given the strategies~$\straa$ and 
$\strab$ (resp., given the strategy~$\straa$).
Given a game $G$ and a state $\l$, a strategy $\straa$ for 
Player~$1$ is \emph{almost-sure winning} 
(resp., \emph{positive winning}) for the objective $\phi$ from $\l$ if 
for all randomized strategies $\strab$ for 
Player~$2$, we have $\Prb_{\l}^{\straa,\strab}(\phi)=1$ 
(resp., $\Prb_{\l}^{\straa,\strab}(\phi)>0$).
Given an MDP $G$ and a state $\l$, a strategy $\straa$ for Player~$1$  
is almost-sure winning (resp. positive winning) for the objective $\phi$ from $\l$
if we have $\Prb_{\l}^{\straa}(\phi)=1$ (resp., $\Prb_{\l}^{\straa}(\phi)>0$).
We also say that state $\l$ is almost-sure winning, or positive winning for $\phi$ respectively.
We are interested in the problems of deciding the existence of an
observation-based strategy for Player~1 
that is almost-sure winning (resp., positive winning) from a given state $\l$.

\begin{comment}

\begin{theorem}[Determinacy] {\rm\cite{Mar75}}
\label{thrm:boreldeterminacy}
For all perfect-information game structures $G$
and all Borel objectives $\phi$,
either there exists a deterministic sure-winning strategy for Player~$1$ for 
the objective $\phi$, or there exists a deterministic sure-winning strategy 
for Player~$2$ for the complementary objective $\Play(G) \setminus \phi$. 
\end{theorem}

Notice that deterministic strategies suffice for sure winning a game:
given a randomized strategy $\straa$ for Player~$1$, let $\straa^D$
be the deterministic strategy such that for all $\rho \in \Pref(G)$, 
the strategy 
$\straa^D(\rho)$ chooses an action from $\supp(\straa(\rho))$. 
Then $\Outcome_1(G,\straa^D) \subseteq \Outcome_1(G,\straa)$, and
thus, if $\straa$ is sure winning, then so is~$\straa^D$.
The result also holds for observation-based strategies
and for perfect-information games.
However, for almost winning, randomized strategies are more powerful 
than deterministic strategies as shown by Example~\ref{ex:example-one}.

\end{comment}

\section{Upper Bounds for the Qualitative Analysis of \POMDP s}
In this section, we present upper bounds for the qualitative analysis of 
\POMDP s.
We first describe the known results.
For qualitative analysis of MDPs, polynomial time upper bounds are known for all 
parity objectives~\cite{luca-thesis,CJH04}.
It follows from the results of~\cite{CDHR07,BBG08} that the decision 
problems for almost-sure winning for \POMDP s with reachability,
safety, and B\"uchi objectives can be solved in EXPTIME. 
It also follows from the results of~\cite{BBG08} that the decision problem 
for almost-sure winning with coB\"uchi objectives and for positive winning with B\"uchi objectives is undecidable if the
strategies are restricted to be pure, and the 
results of~\cite{GIMBERT:2009:HAL-00403463:3} shows that the problem 
remains undecidable even if randomized strategies are considered.
In this section, we complete the results on upper bounds for the 
qualitative analysis of \POMDP s: we present complexity upper bounds for the 
decision problems of positive winning with reachability, safety and 
coB\"uchi objectives. 
The following result for reachability objectives is simple, and 
for a complete and  systematic analysis we present the proof. 

\begin{theorem}\label{thrm_pos_reach}
Given a \POMDP\/ $G$ with a reachability objective and a starting state $\l$,
the problem of deciding whether there is a positive winning strategy from $\l$ 
in $G$ is NLOGSPACE-complete.
\end{theorem}
\begin{myProof} 
The NLOGSPACE-completeness
result for positive reachability for MDPs follows from reductions 
to and from graph reachability.

\smallskip\noindent{\em Reduction to graph reachability.} Given a
\POMDP\/ $G= \tuple{L,\Sigma,\trans,\Obs}$ 
and a set of target states $\target \subseteq L$,
consider the graph $\ov{G}=\tuple{L,E}$ where
$(\l,\l') \in E$ if there exists an action  $\sigma \in \Sigma$ such 
that $\trans(\l,\sigma)(\l')>0$.
Let $\l$ be a starting state, then the following assertions hold:
(a)~if there is a path $\pi$ in $\ov{G}$ from $\l$ to a state $t \in \target$,
then the randomized memoryless strategy for Player~1 in $G$ that plays
all actions uniformly at random ensures that the path $\pi$ is 
executed in $G$ with positive probability (i.e., ensures positive 
winning for $\Reach(\target)$ in $G$ from $\l$); and
(b)~if there is no path in $\ov{G}$ to reach $T$ from $\l$, then 
there is no strategy (and hence no observation-based strategy) for 
Player~1 in $G$ to achieve $\Reach(\target)$.
This shows that positive winning in \POMDP s can be decided in NLOGSPACE.
Graphs are a special case of \POMDP s and hence graph reachability 
can be reduced to reachability with positive probability in \POMDP s,
therefore the problem is NLOGSPACE-complete.
\end{myProof}

\noindent{\bf Positive winning for safety and coB\"uchi objectives.} 
We now show that the decision problem for positive winning 
with safety and coB\"uchi objectives for \POMDP s can be solved in EXPTIME.
We first show with an example that the simple approach of reduction to a 
perfect-information MDP by subset construction and solving the perfect 
information MDP with safety objective for positive winning does not yield 
the desired result.

\begin{example}
Consider the \POMDP\/ shown in \figurename~\ref{fig:subset-pomdp}: in every state
there exists only one action (which we omit for simplicity). In other words,
we have a partially observable Markov chain. 
States $0$, $1$, and $2$ are safe states and form observation $o_1$,
while state $3$ forms observation $o_2$ (which is not in the safe set). 
The state $0$ in $G$ is positive winning for the safety objective as with 
positive probability the state $2$ is reached and then the state $2$ 
is visited forever.
In contrast, consider the perfect information MDP $G^{\K}$ obtained from 
$G$ by subset construction (in this case $G^{\K}$ is a Markov chain).
In $G^{\K}$ from the state $\{1,2\}$, the possible successors are 
$1,2,$ and $3$, and since the observations are different at $1$ and $2$,
as compared to $3$, the successors of $\{1,2\}$ are $\{1,2\}$ and $\{3\}$. 
The reachable set of states in $G^{\K}$ from the state $\{0\}$ is shown 
in \figurename~\ref{fig:subset-pomdp}.
In $G^{\K}$,  the state $\{0\}$ is not positive winning: the state $\{3\}$
is the only recurrent state reachable from $\{0\}$ and hence from the 
state $\{0\}$, with probability~1, the state $\{3\}$ is reached and 
$\{3\}$ is not a safe state. Note that all this holds regardless of the
precise value of nonzero probabilities.
\begin{figure}[t]
 \begin{center}
    \unitlength=.8mm
\def\fsize{\normalsize}

\begin{picture}(166,52)(0,0)

{\fsize

\node[Nmarks=ir, ExtNL=n](x1)(10,18){$0$}
\node[Nmarks=r, ExtNL=n](x2)(38,31){$1$}
\node[Nmarks=r, ExtNL=n](x3)(38,5){$2$}
\node[Nmarks=n, ExtNL=n](x4)(66,31){$3$}

\node[Nmarks=n, Nw=44, Nh=41, Nmr=3, dash={1.5}0, ExtNL=y, NLangle=115, NLdist=2](A1)(24,18){$o_1$}
\node[Nmarks=n, Nw=17, Nh=15, Nmr=3, dash={1.4}0, ExtNL=y, NLangle=270, NLdist=2](A2)(66,31){$o_2$}


\drawloop[ELside=l, ELdist=1, loopCW=y, loopdiam=7, loopangle=90](x2){$\frac{1}{2}$}
\drawloop[ELside=l, ELdist=1, loopCW=y, loopdiam=7, loopangle=90](x3){$1$}
\drawloop[ELside=l, ELdist=1, loopCW=y, loopdiam=7, loopangle=90](x4){$1$}

\drawedge[ELpos=50, ELside=l, ELdist=1, curvedepth=0](x1,x2){$\frac{1}{2}$}
\drawedge[ELpos=50, ELside=r, ELdist=1, curvedepth=0](x1,x3){$\frac{1}{2}$}

\drawedge[ELpos=50, ELside=l, ELdist=1, curvedepth=0](x2,x4){$\frac{1}{2}$}

\node[Nmarks=ir, ExtNL=n](y1)(100,18){$\{0\}$}
\node[Nmarks=r, ExtNL=n, Nadjust=w, Nadjustdist=2](y2)(128,18){$\{1,2\}$}
\node[Nmarks=n, ExtNL=n](y4)(156,18){$\{3\}$}

\drawloop[ELside=l, ELdist=1, loopCW=y, loopdiam=7, loopangle=90](y2){$\frac{1}{2}$}
\drawloop[ELside=l, ELdist=1, loopCW=y, loopdiam=7, loopangle=90](y4){$1$}

\drawedge[ELpos=42, ELside=l, ELdist=1, curvedepth=0](y1,y2){$1$}
\drawedge[ELpos=50, ELside=l, ELdist=1, curvedepth=0](y2,y4){$\frac{1}{2}$}






}
\end{picture}
    \caption{A \POMDP\/ $G$ and the perfect information MDP $G^{\K}$ obtained by subset construction.}
    \label{fig:subset-pomdp}
 \end{center}
\end{figure}
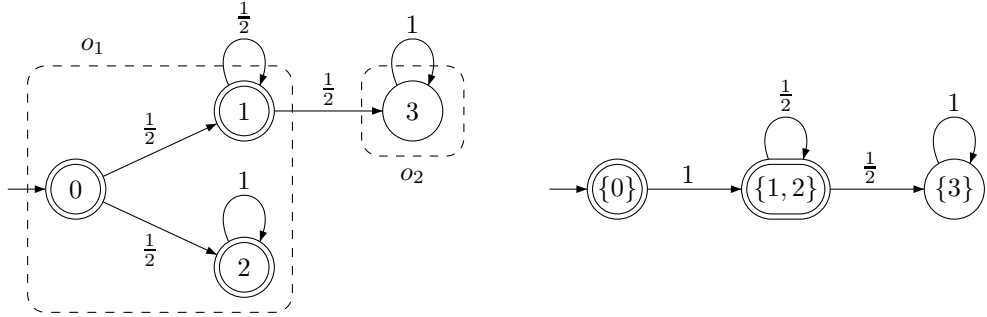
\qed
\end{example}

Our result for positive safety and coB\"uchi objectives is based on the computation 
of almost-sure winning states for safety objectives, and on the following lemma.




\begin{lemma}\label{theo:winningK-then-winning}
Let $G=\tuple{L,\Sigma,\trans,\Obs}$ be a \POMDP\/ and let $\target \subseteq L$ be the set of target states.
If Player~$1$ has an observation-based strategy in $G$ to satisfy $\safe(\target)$ with positive probability
from some state $\l$, then there exists a state $\l'$ such that (a) Player~$1$ has an observation-based strategy 
in $G$ to satisfy $\Until(\target,\{\l'\})$ with positive probability from $\l$, and (b) Player~$1$ has an observation-based 
almost-sure winning strategy in $G$ for $\safe(\target)$ from $\l'$.
\end{lemma}

\begin{myProof}
We assume without loss of generality that the non-safe states in $G$ are absorbing.
Assume that Player~$1$ has an observation-based positive winning strategy $\straa$ in $G$ for the
objective $\safe(\target)$ from~$\l$, and towards a contradiction assume that for all states $\l'$
reachable from $\l$ with positive  probability using $\straa$ in $G$, Player~$1$ has no observation-based 
almost-sure winning strategy for $\safe(\target)$ from~$\l'$. 
A standard argument shows that from every such state $\l'$,
regardless of the observation-based strategy of Player~$1$, the probability to stay safe
within the next $n$ steps is at most $1 - \eta^n$ where $\eta$ is the least non-zero probability in $G$
and $n$ is the number of states in $G$.
Since under strategy $\straa$, every reachable state has this property, the probability to stay 
safe within $k\cdot n$ steps is at most $(1 - \eta^n)^k$. 
This value tends to $0$ when $k \to \infty$, 
therefore the probability to stay safe using $\straa$ from $\l$ is $0$, a contradiction. 
Hence, there exists a state $\l'$ which is almost-sure winning for Player~$1$ (using
observation-based strategy $\straa$) and such that $\l'$ is reached with positive probability 
from $\l$ while staying in $\target$ (again using $\straa$).
\end{myProof}

By Lemma~\ref{theo:winningK-then-winning}, positive winning states can be computed as the
set of states from which Player~$1$ can force with positive probability to reach an almost-sure winning state 
while visiting only safe states. Almost-sure 
winning states can be computed using the following subset construction. 

Given a \POMDP\/ $G=\tuple{L,\Sigma,\trans,\Obs}$ and a set $\target \subseteq L$ of states, 
the \emph{knowledge-based subset construction} of $G$ is the game of perfect observation
\begin{center}
$G^{\K} = \tuple{\L, \Sigma, \trans^{\K}}$, 
\end{center}
where $\L = 2^L \backslash \{\emptyset\}$, and for all $s_1, s_2 \in \L$ (in particular $s_2 \neq \emptyset$)
and $\sigma \in \Sigma$, we have $(s_1, \sigma,s_2) \in \trans^{\K}$ iff there exists an observation
$o \in \Obs$ such that either $s_2 = \Post^G_\sigma(s_1) \cap o \cap \target$,
or $s_2 = (\Post^G_\sigma(s_1) \cap o) \setminus \target$. 
We refer to states in $G^{\K}$ as \emph{cells}.
The following result is established using standard techniques 
(see e.g., Lemma 3.2 and Lemma 3.3 in~\cite{CDHR07}).
and the fact that almost-sure winning and sure winning (sure winning is 
winning with certainty as compared to winning with probability~1 for 
almost-sure winning, see~\cite{CDHR07} for details of sure winning) 
coincide for safety objectives.

\begin{lemma}\label{lem:sure-almost-winning}
Let $G=\tuple{L,\Sigma,\trans,\Obs}$ be a \POMDP\/ and $\target \subseteq L$ a set of target states.
Let $G^{\K}$ be the subset construction and $F_{\target} = \{s \subseteq \target\}$ the set of safe cells.
Player~$1$ has an almost-sure winning observation-based strategy in $G$ for $\safe(\target)$ from $\l$
if and only if  
Player~$1$ has an almost-sure winning strategy in $G^{\K}$ for $\safe(F)$ from cell $\{\l\}$.
\end{lemma}

\begin{remark}
Lemma~\ref{lem:sure-almost-winning} also holds if we replace almost-sure winning
by sure winning, since for safety objectives almost-sure and sure winning 
coincide.
\end{remark}

\begin{theorem}\label{thrm_pos_safe} 
Given a \POMDP\/ $G$ with a safety objective and a starting state $\l$,
the problem of deciding whether there exists a positive winning observation-based 
strategy from $\l$ can be solved in EXPTIME.
\end{theorem}

\begin{myProof}
The almost-sure winning states in $G$ for a safety objective (with observation-based strategy) 
can be computed in exponential time using the subset construction (by Lemma~\ref{lem:sure-almost-winning} and~\cite{CDHR07}).
Then, given the set $W$ of cells that are almost-sure winning in $G^{\K}$, let $\target_W = \{\l \in s \mid s \in W \}$ 
be the almost-sure winning states in $G$. We can compute the states from which Player~$1$
can force $\target_W$ to be reached with positive probability while staying within the safe states
using standard graph analysis algorithms, as in Lemma~\ref{thrm_pos_reach}. Clearly such states
are positive winning in $G$, and by Lemma~\ref{theo:winningK-then-winning} all 
positive winning states in $G$ are obtained in this way. This gives an EXPTIME
algorithm to decide from which states there exists a positive winning observation-based 
strategy for safety objectives.
\end{myProof}

\noindent{\bf Algorithms.}
The complexity bound of Theorem~\ref{thrm_pos_safe} has been established previously in~\cite{GriponS09},
using an extension of the knowledge-based subset construction which is not necessary (where the state space is $L \times 2^L$).
Our proof is simpler and also yield efficient and symbolic algorithms: efficient anti-chain based symbolic 
algorithm for almost-sure winning for safety objectives can be obtained from~\cite{CDHR07}, and positive reachability is 
simple graph reachability. 

The positive winning states for a coB\"uchi objective are computed as
the set of almost-sure winning states for safety that can be reached with positive probability.

\begin{theorem}\label{thrm_pos_cobuchi} 
Given a \POMDP\/ $G$ with a coB\"uchi objective and a starting state $\l$,
the problem of deciding whether there exists a positive winning observation-based strategy from $\l$ 
can be solved in EXPTIME.
\end{theorem}

\begin{myProof}
Let $\coBuchi(\target)$ be a coB\"uchi objective in $G = \tuple{L,\Sigma,\trans,\Obs}$.
As in the proof of Theorem~\ref{thrm_pos_safe}, we compute in exponential time 
the set $\target_W$ of almost-sure winning states in $G$ for $\Safe(\target)$,
and using Lemma~\ref{thrm_pos_reach} the set $W$ of states from which Player~$1$
is positive winning for $\Reach(\target_W)$. Clearly, all states in $W$ are positive
winning for $\coBuchi(\target)$, and $W$ can be computed in EXPTIME. 
We argue that for all states $\l \not \in W$, Player~$1$ is not positive winning
for $\coBuchi(\target)$ from $\l$. Note that $\trans(\l,\sigma)(\l') = 0$
for all $\l \not \in W$, $\l' \in W$, and $\sigma \in \Sigma$, and thus there are no 
almost-sure winning states for $\Safe(\target)$ in $G$ reachable from $L \setminus W$ with positive probability, 
regardless of the strategy of Player~$1$. Therefore, by an argument similar to the proof of 
Lemma~\ref{theo:winningK-then-winning}, for all observation-based strategies for Player~$1$,
from every state $\l \not \in W$, the set $L \setminus \target$ is reached with probability~$1$
and the event $\Buchi(L \setminus \target)$ has probability~$1$. The result follows.
\end{myProof}

\section{Lower Bounds for the Qualitative Analysis of \POMDP s}

In this section we present lower bounds for the qualitative analysis of
\POMDP s. We first present the lower bounds  for MDPs with 
perfect observation.

\smallskip\noindent{\bf Lower bounds for MDPs with perfect observations.}
In the previous section we argued that for reachability objectives even 
in \POMDP s the positive winning problem is NLOGSPACE-complete.
For safety objectives and almost-sure winning it is known that an MDP can 
be equivalently considered as a game where Player~2 makes choices of 
the successors from the support of the probability distribution of the 
transition function, and the almost-sure winning set is the same in the MDP and 
the game. Similarly, there is a reduction of games of perfect observations 
to MDPs of perfect observation for almost-sure winning with safety objectives.
The problem of almost-sure winning in games of perfect observation is
alternating reachability and is PTIME-complete~\cite{Beeri,Immerman81},.
It follows that almost-sure winning for safety objectives in MDPs is 
PTIME-complete.
We now show that the almost-sure winning problem for reachability and 
the positive winning problem for safety objectives is PTIME-complete for
MDPs with perfect observation.

\smallskip\noindent{\bf Reduction from the {\sc Circuit-Value-Problem}.}
Let $N=\set{1,2,\ldots,n}$ be a set of AND and OR gates, 
and $I$ be a set of inputs. 
The set of inputs is partitioned into $I_0$ and $I_1$; $I_0$ is the set of
inputs set to~0 (false) and $I_1$ is the set of inputs set to~1 (true).
Every gate receives two inputs and produces one output; 
the inputs of a gate are outputs of another gate or an input from
the set $I$.
The connection graph of the circuit must be acyclic.
Let the gate represented by the node~1 be the output node. 
The {\sc Circuit-Value-Problem} (CVP) is to decide whether the output is~1 or~0. This problem is PTIME-complete.
We present a reduction of CVP to MDPs with perfect observation for 
almost-sure winning with reachability, and positive winning with safety 
objectives.
\begin{enumerate}
\item \emph{Almost-sure reachability.} Given the CVP, we construct the MDP 
of perfect observation as follows:
(a)~the set of states is $N \cup I$;
(b)~the action set is $\Sigma=\set{l,r}$;
(c)~the transition function is as follows: every node in $I$ is absorbing,
and for a state that represents a gate, (i)~if it is an OR gate, then for the 
action $l$ the left input gate is chosen with probability~1, and for the 
action $r$ the right input gate is chosen with probability~1; and 
(ii)~if it is an AND gate, then irrespective of the action, the left and right 
input gate are chosen with probability~$1/2$.
The output of the CVP from node~1 is~1 iff the set $I_1$ is reached from the state~1
in the MDP with probability~1 (i.e., the state~1 is almost-sure winning for the 
reachability objective $\reach(I_1)$.)

\item \emph{Positive safety.} For positive winning with safety objectives, we 
take the CVP,  apply the same reduction as for almost-sure 
reachability with the following modifications:
every state in $I_0$ remains absorbing and from every state in $I_1$ the next
state is the starting state~1 with probability~1 irrespective of the action.
The set of safety target is the set $I_1 \cup N$. 
If the output of the CVP problem is~1, then from the starting state the set 
$I_1$ is reached with probability~1, and hence the safety objective with the
target $N \cup I_1$ is ensured with probability~1.
If the output of the CVP problem is~0, then from the starting state the set 
$I_0$ is reached with positive probability $\eta>0$ in $n$ steps against 
all strategies. 
Since from every state in $I_1$ the successor state is the state $1$, 
it follows that the probability to reach $I_0$ from the starting state~1 in 
$k\cdot (n+1)$ steps is at least $1-(1-\eta)^k$, and this goes to~1 as 
$k$ goes to $\infty$. 
Hence it follows that from state~1, the answer to the positive winning for
the safety objective $\safe(N \cup I_1)$ is YES iff the output to the CVP 
is~1.
\end{enumerate}
From the above results it also follows that almost-sure and positive B\"uchi and 
coB\"uchi objectives are PTIME-hard (and PTIME-completeness follows from the 
known polynomial time algorithms for qualitative analysis of MDPs with parity 
objectives~\cite{CJH04,luca-thesis}).

\begin{theorem}
Given an MDP $G$ of perfect observation, the following assertions hold:
(a)~the positive winning problem for reachability objectives is NLOGSPACE-complete, and 
the positive winning problem for safety, B\"uchi, coB\"uchi and parity objectives is PTIME-complete; and 
(b)~the almost-sure winning problem for reachability, safety, B\"uchi, coB\"uchi and parity objectives is PTIME-complete. 
\end{theorem}

\noindent{\bf Lower bounds for \POMDP s.}
We have already shown that positive winning with reachability objectives
in \POMDP s is NLOGSPACE-complete.
As in the case of MDPs with perfect observation, for safety objectives and 
almost-sure winning  a \POMDP\/ can be equivalently considered as a game of 
partial observation where Player~2 makes choices of the successors from the 
support of the probability distribution of the transition function, and 
the almost-sure winning set is the same in the \POMDP\/ and the game. 
Since the problem of almost-sure winning in games of partial observation with safety
objective is EXPTIME-complete~\cite{BD08}, the EXPTIME-completeness 
result follows.
We now show that almost-sure winning with reachability objectives and positive 
winning with safety objectives is EXPTIME-complete.
Before the result we first present a discussion on polynomial-space 
alternating Turing machines (ATM).

\noindent{\em Discussion.} Let $M$ be a polynomial-space ATM and let $w$ be an input word.
Then, there is an exponential bound on the number of configurations of the machine. 
Hence if $M$ can accept the word $w$, then it can do so within some $k_{\abs{w}}$ steps, 
where $\abs{w}$ is the length of the word $w$, and $k_{\abs{w}}$ is bounded by an exponential in $\abs{w}$. 
We construct an equivalent polynomial-space ATM $M'$ that behaves as $M$ but keeps track (in polynomial space) 
of the number of steps executed by $M$, and given a word $\abs{w}$, 
if the number of steps reaches $k_{\abs{w}}$ without accepting, then the word is rejected.
The machine $M'$ is equivalent to $M$ and reaches the accepting or rejecting states in a number
of steps bounded by an exponential in the length of the input word.
The problem of deciding, given a polynomial-space ATM $M$ and a word $w$, whether $M$ accepts $w$ is 
EXPTIME-complete.

\smallskip\noindent{\bf Reduction from Alternating PSPACE Turing machine.} 
Let $M$ be a polynomial-space ATM such that for every input word $w$,
the accepting or the rejecting state is reached within exponential steps in $\abs{w}$.
A polynomial-time reduction $R_G$ of a polynomial-space ATM $M$ and an input word $w$ to a 
game $G = R_G(M,w)$ of partial observation is given in~\cite{CDHR07} such that 
(a)~there is a special accepting state in $G$, and
(b)~$M$ accepts $w$ iff there is an observation-based strategy for Player~1 in $G$ 
to reach the accepting state with probability~1.
If the above reduction is applied to $M$, then the game structure satisfies the 
following additional properties: there is a special rejecting state that is absorbing, 
and for every observation-based strategy for Player~1, either 
(a)~against all Player~2 strategies the 
accepting state is reached with probability~1; or (b)~there is a pure Player~2 
strategy that reaches the rejecting state with positive probability $\eta>0$ in 
$2^{\abs{L}}$ steps and the accepting or the rejecting state is reached with 
probability~1 in $2^{\abs{L}}$ steps.
We now present the reduction to \POMDP s:
\begin{enumerate}

\item \emph{Almost-sure winning for reachability.}
Given a polynomial-space ATM $M$ and $w$ an input word, let $G=R_G(M,w)$. We construct a \POMDP\/ $G'$ from 
$G$ as follows: we only modify the transition function in $G'$ by uniformly choosing over
the successor choices. 
Formally, for a state $\l \in L$ and an action $\sigma \in \Sigma$ 
the probabilistic transition function $\trans'$ in $G'$ is as follows:
\[
\trans'(\l,\sigma)(\l') = 
\begin{cases}
0 & (\l,\sigma,\l') \not \in \trans; \\
1/ \abs{\set{\l_1 \mid (\l,\sigma,\l_1) \in \trans}} & (\l,\sigma,\l') \in \trans.
\end{cases}
\]
Given an observation-based strategy for Player~1 in $G$, we consider the same
strategy in $G'$: 
(1)~if the strategy reaches the accepting state with probability~1
against all Player~2 strategies in $G$, then the strategy ensures that in $G'$
the accepting state is reached with probability~1; and
(2)~otherwise there is a pure Player~2 strategy $\strab$ in $G$ that ensures the rejecting
state is reached in $2^{\abs{L}}$ steps with probability $\eta>0$, and with 
probability at least $(1/\abs{L})^{2^{\abs{L}}}$ the choices of the successors of strategy $\strab$ 
is chosen in $G'$, and hence the rejecting state is reached with probability at least
$(1/\abs{L})^{2^{\abs{L}}} \cdot \eta>0$. 
It follows that in $G'$ there is an observation-based strategy for almost-sure winning 
the reachability objective with target of the accepting state iff there is such a 
strategy in $G$.
The result follows.

\item \emph{Positive winning for safety.} 
The reduction is same as above. We obtain the \POMDP\/ $G''$ from the \POMDP\/ $G'$ above 
by making the following modification: from the state accepting, the \POMDP\/ goes back to the 
initial state with probability~1.
If there is an observation-based strategy $\straa$ for Player~1 in $G'$ to reach the 
accepting state, then repeating the strategy $\straa$ each time the accepting state is 
visited, it can be ensured that the rejecting state is reached with probability~0.
Otherwise, against every observation-based strategy for Player~1, the probability to 
reach  the rejecting state in $k \cdot(2^{\abs{L}}+1)$ steps is at least 
$1-(1-\eta')^k$, where $\eta'=\eta \cdot (1/\abs{L})^{2^{\abs{L}}}>0$ (this is 
because there is a probability to reach the rejecting state with probability at least 
$\eta'$ in $2^{\abs{L}}$ steps, and unless the rejecting state is reached the starting state 
is again reached within $2^{\abs{L}}+1$ steps).
Hence the probability to reach the rejecting state is~1.
It follows that $G'$ is almost-sure winning for the reachability objective with the 
target of the accepting state iff in $G''$ there is an observation-based strategy 
for Player~1 to ensure that the rejecting state is avoided with positive probability.
This completes the proof of correctness of the reduction.
\end{enumerate}

A very brief (two line proof) sketch was presented as the proof of Theorem~1
of~\cite{dA99} to show that positive winning in \POMDP s with safety objectives
is EXPTIME-hard. 
We were unable to reconstruct the proof: the proof suggested to simulate a 
nondeterministic Turing machine. 
The simulation of a polynomial-space nondeterministic Turing machine only shows 
PSPACE-hardness, and the simulation of a nondeterministic EXPTIME Turing 
machine would have shown NEXPTIME-hardness, and an EXPTIME upper bound is known
for the problem. Our proof presents a different and detailed proof of the 
result of Theorem~1 of~\cite{dA99}.
Hence we have the following theorem, and the results are summarized in Table~\ref{tab1}. 

\begin{theorem}\label{thrm_complexity}
Given a \POMDP\ $G$, the following assertions hold:
(a)~the positive winning problem for reachability objectives is NLOGSPACE-complete,
the positive winning problem for safety and coB\"uchi objectives is EXPTIME-complete, 
and the positive winning problem for B\"uchi and parity objectives is undecidable; and
(b)~the almost-sure winning problem for reachability, safety and B\"uchi objectives is 
EXPTIME-complete, and the almost-sure winning problem for coB\"uchi and parity objectives is 
undecidable.
\end{theorem}
\begin{myProof} 
The results are obtained as follows.
\begin{enumerate}
\item {\em Positive winning.}
The NLOGSPACE-completeness for positive winning with reachability objectives is 
Theorem~\ref{thrm_pos_reach}.
Our reduction from Alternating PSPACE Turing machine shows EXPTIME-hardness for positive 
winning with safety (and hence the lower bound also follows for coB\"uchi objectives), 
and the upper bounds follow from Theorem~\ref{thrm_pos_safe} and Theorem~\ref{thrm_pos_cobuchi}.
The undecidability follows for positive winning for B\"uchi and parity objectives
follows from the result of~\cite{BBG08,GIMBERT:2009:HAL-00403463:3}.

\item {\em Almost-sure winning.} 
It follows from the results of~\cite{CDHR07,BBG08} that the decision 
problems for almost-sure winning for \POMDP s with reachability,
safety, and B\"uchi objectives can be solved in EXPTIME.
Our reduction from Alternating PSPACE Turing machine shows EXPTIME-hardness for almost-sure
winning with reachability (and hence the lower bound also follows for B\"uchi objectives). 
The lower bound for safety objectives follows from the lower bound for partial information 
games~\cite{CDHR07} and the fact the almost-sure winning for safety coincides 
with almost-sure winning in games.
The undecidability follows for almost-sure winning for coB\"uchi and parity objectives
follows from the result of~\cite{BBG08,GIMBERT:2009:HAL-00403463:3}.
\end{enumerate}
\end{myProof}

\begin{table}
\begin{center}
\begin{tabular}{|c|c|c|}
\hline
             & Positive &  Almost-sure \\
\hline
\,Reachability\, & \,NLOGSPACE-complete (up+lo)\, & EXPTIME-complete (lo)\\ 
\hline
Safety       & EXPTIME-complete (up+lo) &  \,EXPTIME-complete\,~\cite{BD08}\, \\ 
\hline
B\"uchi      & Undecidable~\cite{BBG08} & EXPTIME-complete (lo) \\ 
\hline
coB\"uchi     & EXPTIME-complete (up+lo) & Undecidable~\cite{BBG08} \\ 
\hline
Parity        & Undecidable~\cite{BBG08} & Undecidable~\cite{BBG08} \\ 
\hline
\hline
\end{tabular}
\end{center}
\caption{Computational complexity of \POMDP s with different classes of 
parity objectives for positive and almost-sure winning. Our contribution of 
upper and lower bounds are indicated as ``up'' and ``lo'' respectively 
in parenthesis.}\label{tab1}
\end{table}

\section{Optimal Memory Bounds for Strategies}
In this section we present optimal bounds on the memory required by 
pure and randomized strategies for positive and almost-sure winning for 
reachability, safety, B\"uchi and coB\"uchi objectives.

\smallskip\noindent{\bf Bounds for safety objectives.}
First, we consider positive and almost-sure winning with 
safety objectives in \POMDP s. 
It follows from the correctness argument of Theorem~\ref{thrm_pos_safe} 
that pure strategies 
with exponential memory are sufficient for positive winning with 
safety objectives in \POMDP s, and the exponential upper bound on memory
of pure strategies for almost-sure winning with safety objectives in \POMDP s
follows from the reduction to games.
We now present a matching exponential lower bound for randomized strategies.

\begin{lemma}\label{lemm_lower_bound_safety}
There exists a family $(P_n)_{n \in \nat}$ of \POMDP s 
of size $O(p(n))$ for a polynomial $p$ with a safety 
objective  such that the following assertions hold:
(a)~Player~$1$ has a (pure) almost-sure (and therefore also positive)
winning strategy in each of these \POMDP s; and 
(b)~there exists a polynomial $q$ such that 
every finite-memory randomized strategy for Player~1 that  is 
positive (or almost-sure) winning 
in $P_n$ has at least $2^{q(n)}$ states.
\end{lemma}

\paragraph{{\bf Preliminary.}}
The set of actions of the \POMDP\/ $P_n$ is $\Sigma_n \cup \{\#\}$
where $\Sigma_n = \{1, \dots,n\}$. 
The \POMDP\/ is composed of an initial state $q_0$ and $n$ sub-MDPs $A_i$ with state space $Q_i$, 
each consisting  of a loop over $p_i$ states $q_1^i,\dots,q_{p_i}^i$
where $p_i$ is the $i$-th prime number. 
From each state $q_j^i$ ($1 \leq j < p_i$), every action in $\Sigma_n$ 
leads to the next state $q_{j+1}^i$ with probability $\frac{1}{2}$, and to the initial state $q_0$ 
with probability $\frac{1}{2}$. The action $\#$ is not allowed.
From $q_{p_i}^i$, the action $i$ is not allowed
while the other actions in $\Sigma_n$ lead back the first state $q^i_1$
and to the initial state $q_0$ both with probability $\frac{1}{2}$.
Moreover, the action $\#$ leads back to the initial state (with probability~$1$). 
The disallowed actions lead to a bad state. The states of the $A_i$'s are 
indistinguishable (they have the same observation), while the initial state $q_0$ is visible.
We assume that the state spaces $Q_i$ of the $A_i$'s are disjoint.

\begin{figure}[t]
 \begin{minipage}[b]{.52\linewidth}
 \begin{center}
    \unitlength=.8mm
\def\fsize{\normalsize}

\scalebox{.75}{
\begin{picture}(110,80)(0,0)

\put(5,0){
{\fsize

\node[Nmarks=i, iangle=90](q0)(45,65){$q_0$}

\node[Nmarks=n](q1)(20,35){$q^1_1$}
\node[Nmarks=n](q2)(20,15){$q^1_2$}

\node[Nmarks=n](q3)(70,35){$q^2_1$}
\node[Nmarks=n](q4)(60,15){$q^2_2$}
\node[Nmarks=n](q5)(80,15){$q^2_3$}


\node[Nmarks=n, Nw=31, Nh=35, dash={1.5}0, ExtNL=y, NLangle=115, NLdist=2](A1)(21,25){$A_1$}
\node[Nmarks=n, Nw=41, Nh=41, dash={1.5}0, ExtNL=y, NLangle=55, NLdist=2](A2)(70,22){$A_2$}

\drawedge[ELpos=60, ELside=l, ELdist=1, sxo=7, exo=-2, curvedepth=-6](A1,q0){$\frac{1}{2}$}
\drawedge[ELpos=65, ELside=r, ELdist=1, sxo=-10, syo=-3, exo=1, curvedepth=6](A2,q0){$\frac{1}{2}$}


\drawedge[ELpos=50, ELside=r, ELdist=1, curvedepth=-6](q0,q1){$\frac{1}{2}$}
\drawedge[ELpos=50, ELside=l, ELdist=1, curvedepth=6](q0,q3){$\frac{1}{2}$}

\drawedge[ELpos=50, ELside=r, ELdist=1, curvedepth=-5](q1,q2){$\Sigma_2$}
\drawedge[ELpos=50, ELside=r, ELdist=1, curvedepth=-5](q2,q1){$\Sigma_2 \!\setminus\! \{1\}$}
\drawedge[ELpos=50, ELside=l, ELdist=1, curvedepth=-5](q1,q2){$\frac{1}{2}$}
\drawedge[ELpos=50, ELside=l, ELdist=1, curvedepth=-5](q2,q1){$\frac{1}{2}$}

\drawedge[ELpos=50, ELside=r, ELdist=1, curvedepth=-4](q3,q4){$\Sigma_2$}
\drawedge[ELpos=50, ELside=l, ELdist=1.5, curvedepth=-4](q4,q5){$\Sigma_2$}
\drawedge[ELpos=40, ELside=r, ELdist=1, curvedepth=-4](q5,q3){$\Sigma_2 \!\setminus\! \{2\}$}

\drawedge[ELpos=50, ELside=l, ELdist=1, curvedepth=-4](q3,q4){$\frac{1}{2}$}
\drawedge[ELpos=50, ELside=r, ELdist=1.5, curvedepth=-4](q4,q5){$\frac{1}{2}$}
\drawedge[ELpos=50, ELside=l, ELdist=1, curvedepth=-4](q5,q3){$\frac{1}{2}$}



\node[Nframe=n, Nw=0,Nh=0](tmp1)(0,17){}
\node[Nframe=n, Nw=0,Nh=0](tmp2)(0,63){}
\drawline[AHnb=0, arcradius=2](15,15)(0,15)(0,25)
\drawedge[AHnb=0, ELpos=50, ELside=l, ELdist=1, curvedepth=0](tmp1,tmp2){$\#$}
\drawline[AHnb=1, arcradius=2](0,45)(0,65)(40,65)

\node[Nframe=n, Nw=0,Nh=0](tmp1)(96,17){}
\node[Nframe=n, Nw=0,Nh=0](tmp2)(96,63){}
\drawline[AHnb=0, arcradius=2](85,15)(96,15)(96,25)
\drawedge[AHnb=0, ELpos=50, ELside=r, ELdist=1, curvedepth=0](tmp1,tmp2){$\#$}
\drawline[AHnb=1, arcradius=2](96,45)(96,65)(50,65)





}}
\end{picture}
}
    \caption{The \POMDP\/ $P_2$.}
    \label{fig:exp-game}
 \end{center}
 \end{minipage} \hfill 
 \begin{minipage}[b]{.52\linewidth}
 \begin{center}
   \unitlength=.8mm
\def\fsize{\normalsize}

\scalebox{.75}{
\begin{picture}(90,90)(0,0)

{\fsize

\node[Nmarks=i, iangle=90](q0)(35,77){$q_0$}
\node[Nmarks=n](q1)(15,57){$q^1_1$}
\node[Nmarks=n](q2)(15,37){$q^1_2$}

\node[Nmarks=n](q3)(55,57){$q^2_1$}
\node[Nmarks=n](q4)(45,37){$q^2_2$}
\node[Nmarks=n](q5)(65,37){$q^2_3$}

\node[Nmarks=n, Nadjust=wh, Nadjustdist=2](safe)(30,17){$\Goal$}

\node[Nmarks=n, Nw=34, Nh=35, dash={1.5}0, ExtNL=y, NLangle=115, NLdist=2](A1)(19,47){$H_1$}
\node[Nmarks=n, Nw=45, Nh=35, dash={1.5}0, ExtNL=y, NLangle=55, NLdist=2](A2)(61,47){$H_2$}


\drawedge[ELpos=50, ELside=r, ELdist=1, curvedepth=0](q0,q1){$\frac{1}{2}$}
\drawedge[ELpos=50, ELside=l, ELdist=1, curvedepth=0](q0,q3){$\frac{1}{2}$}

\drawedge[ELpos=50, ELside=r, ELdist=1, curvedepth=-4](q1,q2){$\tick$}
\drawedge[ELpos=50, ELside=r, ELdist=1, curvedepth=-4](q2,q1){$\tick$}

\drawedge[ELpos=50, ELside=r, ELdist=1, curvedepth=-4](q3,q4){$\tick$}
\drawedge[ELpos=50, ELside=l, ELdist=1.5, curvedepth=-4](q4,q5){$\tick$}
\drawedge[ELpos=40, ELside=r, ELdist=1, curvedepth=-4](q5,q3){$\tick$}

\drawedge[ELpos=50, ELside=r, ELdist=1, curvedepth=0](q2,safe){$\#$}
\drawedge[ELpos=50, ELside=l, ELdist=1, curvedepth=0, syo=-3](q5,safe){$\#$}

\drawloop[ELside=r,loopCW=n, loopdiam=6, loopangle=-90](safe){$\#$}



}
\end{picture}
}
   \caption{The \POMDP\/ $P'_2$.}
   \label{fig:exp-POMDP-reach}
 \end{center}
 \end{minipage}
\end{figure}

\paragraph{{\bf \POMDP\/ family $(P_n)_{n \in\nat}$.}}
The state space of $P_n$ is the disjoint union of $Q_1, \dots, Q_n$ and $\{q_0,\Bad\}$.
The initial state is $q_0$, the final state is $\Bad$. The probabilistic 
transition function is as follows:
\begin{itemize}
\item for all $1 \leq i \leq n$ and $\sigma \in \Sigma_n$, we have
$\trans(q_0,\sigma)(q^i_1)=\frac{1}{n}$;

\item 
for all $1 \leq i \leq n$, $1 \leq j < p_i$, and $\sigma \in \Sigma_n$, 
$\sigma' \in \Sigma_n \setminus \{i\}$,
we have 
$\trans(q^i_j,\sigma)(q^i_{j+1})=\trans(q^i_{j},\sigma)(q_{0})= 
\trans(q^i_{p_i},\sigma')(q^i_{1})=\trans(q^i_{p_i},\sigma')(q_{0})=\frac{1}{2}$;
and 

\item for all $1 \leq i \leq n$ and $1 \leq j < p_i$, we have 
$\trans(q_0,\#)(\Bad)=\trans(q^i_{j},\#)(\Bad)=\trans(q^i_{p_i},\#)(q_{0})=1$. 
\end{itemize}

The initial state is $q_0$. There are two observations, the state $\{q_0\}$ is 
labelled by observation $o_1$, 
and the other states in $Q_1 \cup \dots \cup Q_n$ (that we call the loops) 
by observation $o_2$.
\figurename~\ref{fig:exp-game} shows the game $P_2$: 
the witness family of POMDPs have similarities with analogous constructions for games~\cite{BCDHR08}.
However the construction of~\cite{BCDHR08} shows lower bounds only for pure strategies and in games, whereas
we present lower bound for randomized strategies and for POMDPs, and hence our proofs are very different.

\paragraph{{\bf Proof of Lemma~\ref{lemm_lower_bound_safety}.}}
After the first transition from the initial state, player~$1$ 
has the following positive winning strategy. Let $p^*_n = \prod_{i=1}^{n} p_i$. 
While the \POMDP\/ is in the loops (assume that we have seen $j$ times observation $o_2$
consecutively), if $1 \leq j < p^*_n$, then play any action $i$ such that $j \mod p_i \neq 0$
(this is well defined since $p^*_n$ is the lcm of $p_1, \dots, p_n$),
and otherwise play $\#$. It is easy to show that this strategy is winning for 
the safety condition, with probability $1$.

For the second part of the result, assume towards a contradiction 
that there exists a finite-memory randomized strategy $\hat{\straa}$ 
that is positive winning 
for Player~$1$ and has less than $p^*_n$ states 
(since $p^*_n$ is exponential in $s^*_n = \sum_{i=1}^{n} p_i$, 
the result will follow).
Let $\eta$ be the least positive transition probability described by 
the finite-state strategy $\hat{\straa}$. 
Consider any history of a play $\rho$ that ends with $o_1$. 
We claim that the following properties hold:
(a)~with probability~$1$ either observation $o_1$ is visited again from $\rho$ or 
the state $\Bad$ is reached; and 
(b)~the state $\Bad$ is reached with a positive probability.
The first property (property~(a)) follows from the fact that for all actions
the loops are left (the state $q_0$ or $\Bad$ is reached) with probability at 
least $\frac{1}{2}$.
We now prove the second property by showing that the state~$\Bad$ is reached with 
probability at least $\Delta_n = \frac{1}{n}\cdot\frac{1}{(2\cdot \eta)^{p^*_n}}$.
To see this, consider the sequence of actions played by strategy $\hat{\straa}$
after~$\rho$ when only~$o_2$ is observed. Either $\#$ is never played, and then
the action played by $\hat{\straa}$ after a sequence of $p^*_n$ states leads 
to $\Bad$ (the current state being then $q^i_{p_i}$ for some $1 \leq i \leq n$).
This occurs with probability at least~$\Delta_n$; or~$\#$ is eventually played,
but since $\hat{\straa}$ has less than $p^*_n$ states, it has to be played
after less than $p^*_n$ steps, which also leads to $\Bad$ with probability at 
least~$\Delta_n$. 
The above two properties that (a)~$o_1 \cup \{\Bad\}$ is reached with probability 
$1$ from $o_1$, 
and (b)~within $p^*_n$ steps after a visit to $o_1$, the state $\Bad$ is reached with
fixed positive probability,
ensures that $\Bad$ is reached with probability~$1$.
Hence $\hat{\straa}$ is not positive winning.
It follows that randomized strategies that are almost-sure or positive winning in 
\POMDP s with safety objectives may require exponential memory.

\smallskip\noindent{\bf Bounds for reachability objectives.}
We now argue the memory bounds
for pure and randomized strategies for positive winning with reachability 
objectives.
\begin{enumerate}

\item It follows from the correctness argument of Theorem~\ref{thrm_pos_reach} 
that randomized memoryless strategies suffice for positive winning with reachability 
objectives in \POMDP s.

\item We now argue that for pure strategies, memory of size linear in the number of 
states is sufficient and may be necessary. 
The upper bound follows from the reduction to graph reachability.
Given a \POMDP\/ $G$, consider the graph $\ov{G}$ constructed from $G$ as 
in the correctness argument for Theorem~\ref{thrm_pos_reach}. 
Given the starting state $\l$, if there is path in $\ov{G}$ to the target set 
$T$ obtained from $\target$, then there is a path $\pi$ of length at most $\abs{L}$. 
The pure strategy for Player~1 in $G$ can play the sequence of actions
of the path $\pi$ to ensure that the target observations $\target$ 
are reached with positive probability in $G$.
The family of examples to show that pure strategies require linear memory 
can be constructed as follows: we construct a \POMDP\/ with deterministic
transition function such that there is a unique path (sequence of actions)
of length $O(\abs{L})$ to the target, and any deviation leads to an absorbing 
state, and other than the target state every other state has the same 
observation. In this \POMDP\/ any pure strategy must remember the exact 
sequence of actions to be played and hence requires $O(\abs{L})$ memory.
\end{enumerate}
It follows from the results of~\cite{BBG08} that for almost-sure winning 
with reachability
objectives in \POMDP s pure strategies with exponential memory suffice, and 
we now prove an exponential lower bound for randomized strategies.

\begin{lemma}\label{lemm_lower_bound_reachability}
There exists a family $(P_n)_{n \in \nat}$ of \POMDP s 
of size $O(p(n))$ for a polynomial $p$ with a reachability 
objective  such that the following assertions hold:
(a)~Player~$1$ has an almost-sure 
winning strategy in each of these \POMDP s; and 
(b)~there exists a polynomial $q$ such that 
every finite-memory randomized strategy for Player~1 that  is 
almost-sure winning 
in $P_n$ has at least $2^{q(n)}$ states.
\end{lemma}

Fix the action set as  $\Sigma = \{\#,\tick\}$. 
The \POMDP\/ $P'_n$ is composed of an initial state $q_0$ and $n$ sub-MDPs $H_i$, 
each consisting of a loop over $p_i$ states $q_1^i,\dots,q_{p_i}^i$ where $p_i$ is the $i$-th prime number.
From each state in the loops, the action $\tick$ can be played and leads to 
the next state in the loop (with probability $1$). The action $\#$ can be played 
in the last state of each loop and leads to the $\Goal$ state. The objective
is to reach $\Goal$ with probability~1. Actions that are not allowed lead to a sink
state from which it is impossible to reach $\Goal$.
There is a unique observation that consists of the whole state space.
\figurename~\ref{fig:exp-POMDP-reach} shows $P'_2$.

\paragraph{{\bf Proof of Lemma~\ref{lemm_lower_bound_reachability}.}}
First we show that Player $1$ has an almost-sure winning 
strategy in $P'_k$ (from $q_0$). 
As there is only one observation,
a strategy for Player $1$ corresponds to a function 
$\straa: \nat \to \Sigma$.
Consider the strategy $\straa^*$ as follows:
$\straa^*(j) = \tick$ for all $0 \leq j < p^*_k$
and $\straa^*(j) = \#$ for all $j \geq p^*_k$.
It is easy to check that $\straa^*$ ensures winning 
with certainty and hence almost-sure winning.

For the second part of the result  assume, towards a contradiction, 
that there exists a finite-memory randomized strategy 
$\hat{\straa}$ 
that is almost-sure winning and has less than $p^*_k$ states. 
Clearly, 
 $\hat{\straa}$ cannot play $\#$ before the $(p^*_k+1)$-th round
since one of the subMDPs $H_i$ would 
not be in $q^i_{p_i}$ and therefore Player $1$ would lose with probability at least $\frac{1}{n}$. 
Note that the state reached 
by the strategy automaton defining $\hat{\straa}$ after $p^*_k$ rounds has necessarily
been visited in a previous round. Since $\hat{\straa}$ has to play $\#$ eventually to reach $\Goal$,
this means that $\#$ must have been played in some round $j < p^*_k$, 
when at least one of the subgames~$H_i$ was not in location $q^i_{p_i}$, 
so that Player~$1$ would have already lost with probability at least $\frac{1}{n}\cdot \eta$,
where $\eta$ is the least positive probability specified by $\hat{\straa}$.
This is in contradiction with our assumption that $\hat{\straa}$ is an almost-sure winning strategy.

\medskip\noindent{\bf Bounds for B\"uchi and coB\"uchi objectives.} 
An exponential upper bound for memory of pure strategies for almost-sure winning of 
B\"uchi objectives follows from the results of~\cite{BBG08}, and the matching lower
bound for randomized strategies follows from our result for reachability 
objectives.
Since positive winning is undecidable for B\"uchi objectives there is no bound 
on memory for pure or randomized strategies for positive winning.
An exponential upper bound for memory of pure strategies for positive winning 
of coB\"uchi objectives follows from the correctness proof of 
Theorem~\ref{thrm_pos_cobuchi} that iteratively combines the positive winning 
strategies for safety and reachability to obtain a positive winning strategy 
for coB\"uchi objective. 
The matching lower bound for randomized strategies follows from our result for 
safety objectives.
Since almost-sure winning is undecidable for coB\"uchi objectives there is no bound 
on memory for pure or randomized strategies for positive winning.
This gives us the following theorem (also summarized in Table~\ref{tab2}), 
which is in contrast to the results for MDPs with perfect observation where 
pure memoryless strategies suffice for almost-sure and positive winning for all 
parity objectives.

\begin{theorem}
The optimal memory bounds for strategies in \POMDP s are as follows.
\begin{enumerate}
\item Reachability objectives: for positive winning randomized memoryless 
strategies are sufficient, and linear memory is necessary and sufficient for pure 
strategies; and for almost-sure winning exponential memory is necessary and 
sufficient for both pure and randomized strategies.  

\item Safety objectives: for positive winning and almost-sure winning
exponential memory is necessary and sufficient for both pure and 
randomized strategies.  

\item B\"uchi objectives: for almost-sure winning exponential memory is necessary 
and sufficient for both pure and randomized strategies; and there is no bound 
on memory for pure and randomized strategies for positive winning.

\item coB\"uchi objectives: for positive winning exponential memory is 
necessary and sufficient for both pure and randomized strategies; and there 
is no bound on memory for pure and randomized strategies for almost-sure winning.
\end{enumerate}

\end{theorem}

\begin{table}
\begin{center}
\begin{tabular}{|c|c|c|c|c|}
\hline
             & Pure Positive & Randomized Positive   & Pure Almost  & Randomized Almost \\
\hline
Reachability & Linear & Memoryless & Exponential & Exponential \\ 
\hline
Safety       & Exponential & Exponential & Exponential & Exponential \\ 
\hline
B\"uchi       & No Bound & No Bound & Exponential & Exponential \\ 
\hline
coB\"uchi     & Exponential & Exponential & No Bound & No Bound \\ 
\hline
Parity        & No Bound & No Bound & No Bound & No Bound \\ 
\hline
\hline
\end{tabular}
\end{center}
\caption{Optimal memory bounds for pure and randomized strategies for positive and
almost-sure winning.}\label{tab2}
\end{table}

\bibliographystyle{plain}
\bibliography{diss}

\end{document}